\documentclass[
  aps,              
  pra,              
   preprint,       
  showkeys,         
  groupedaddress,   
superscriptaddress, 
   floatfix         
  ]{revtex4-2}
  
\usepackage{amsmath}
\usepackage{amssymb}
\usepackage{fourier}

\usepackage{graphicx,dcolumn,bm,algpseudocode,mathtools,cases,float}

\usepackage[utf8]{inputenc}
\usepackage[T1]{fontenc}

\usepackage[hidelinks]{hyperref}
\hypersetup{
  colorlinks=true,
  allcolors=blue,
  urlcolor=blue
  }

\usepackage[dvipsnames]{xcolor}
\usepackage{totcount}
\newcounter{tofixn}

\regtotcounter{tofixn}

\begin{document}

\title[Zhang, et al.: Capillary waves from acoustic excitation]{Onset of visible capillary waves from high-frequency acoustic excitation}

\author{Shuai Zhang}
\affiliation{Medically Advanced Devices Laboratory, Center for Medical Device Engineering and Biomechanics, Department of Mechanical and Aerospace Engineering, Jacobs School of Engineering, University of California San Diego, La Jolla, CA 92093-0411 USA}
\affiliation{Materials Science and Engineering Program, Jacobs School of Engineering, University of California San Diego, La Jolla, CA 92093-0411 USA}
\author{Jeremy Orosco}
\affiliation{Medically Advanced Devices Laboratory, Center for Medical Device Engineering and Biomechanics, Department of Mechanical and Aerospace Engineering, Jacobs School of Engineering, University of California San Diego, La Jolla, CA 92093-0411 USA}
\author{James Friend}
\email[Corresponding author: ]{jfriend@ucsd.edu}
\affiliation{Medically Advanced Devices Laboratory, Center for Medical Device Engineering and Biomechanics, Department of Mechanical and Aerospace Engineering, Jacobs School of Engineering, University of California San Diego, La Jolla, CA 92093-0411 USA}
\affiliation{Materials Science and Engineering Program, Jacobs School of Engineering, University of California San Diego, La Jolla, CA 92093-0411 USA}
\affiliation{Department of Surgery, School of Medicine,     University of California San Diego, La Jolla, CA 92093 USA}

\date{\today}

\begin{abstract}


Remarkably, the interface of a fluid droplet will produce visible capillary waves when exposed to acoustic waves. For example, a small ($\sim\!1\, \mu$L) sessile droplet will oscillate at a low $\sim\!10^2$~Hz frequency when weakly driven by acoustic waves at $\sim\!10^6$~Hz frequency and beyond. We measured such a droplet's interfacial response to 6.6~MHz ultrasound to gain insight into the energy transfer mechanism that spans these vastly different timescales, using high-speed microscopic digital transmission holography, a unique method to capture three-dimensional surface dynamics at nanometer space and microsecond time resolutions. We show that low-frequency capillary waves are driven into existence via a feedback mechanism between the acoustic radiation pressure and the evolving shape of the fluid interface. The acoustic pressure is distributed in the standing wave cavity of the droplet, and as the shape of the fluid interface changes in response to the distributed pressure present on the interface, the standing wave field also changes shape, feeding back to produce changes in the acoustic radiation pressure distribution in the cavity. A physical model explicitly based upon this proposed mechanism is provided, and simulations using it were verified against direct observations of both the microscale droplet interface dynamics from holography and internal pressure distributions using microparticle image velocimetry. The pressure-interface feedback model accurately predicts the vibration amplitude threshold at which capillary waves appear, the subsequent amplitude and frequency of the capillary waves, and the distribution of the standing wave pressure field within the sessile droplet responsible for the capillary waves.
\end{abstract}

\maketitle


\section{\label{sec:level1}Introduction}

    High-frequency acoustic waves at and beyond 1~MHz is useful in droplet manipulation, fluid mixing, and atomization~\cite{ang2015nozzleless,collignon2018improving}, among many other micro and nanofluidic applications---a relatively new discipline called \emph{acoustofluidics} \cite{Rufo:2022vq,Friend:2011ss}. The challenges of overcoming surface and viscous-dominated phenomena at these scales has been the principal motivation in the development of this field, where the acoustic wave behavior is at spatiotemporal scales commensurate with these applications. For example, acoustic waves at high frequencies may drive atomization from a fluid interface. Capillary waves appear on the free interface~\cite{rayleigh1879capillary} and begin ejecting small droplets from their crests~\cite{lang1962ultrasonic}. Ultrasonic nebulizers offer several advantages over mechanical atomizers and jet nebulizers, including improved portability, narrow droplet size distributions (when properly controlled), good efficiency, and ease of use. Ultrasonic nebulizers are widely used in pulmonary drug delivery~\cite{taylor1997ultrasonic,Rajapaksa:2014uo}, surface coating~\cite{majumder2010insights}, and many other fields. 

    The phenomenon of driving capillary waves on a droplet's surface from vibration has consistently received attention over the years \cite{miles1993faraday,marston1980quadrupole,ockendon1973resonant,benjamin1954stability,rayleigh1879capillary}. Many have studied the droplet's behavior due to exposure to low-frequency vibrations~\cite{whitehill2010droplet}, even looking at the broader spectral response to look for subharmonics \cite{keolian1981subharmonic} and intermittency \cite{craik1995faraday}, hallmarks of nonlinearity. In those cases where ultrasound has been used, it has generally been modulated near the droplet's resonance frequency~\cite{trinh1982experimental,baudoin2012low}: the high-frequency ultrasound serves as a pseudo-static acoustic pressure source. Moreover, at the relatively low forcing frequencies used in classic studies, capillary wave generation has been successfully explained by classical Faraday instability theory~\cite{kumar1996linear} and closely related methods~\cite{murray1999droplet,lyubimov2006behavior}. 
    
    However, the frequencies typically used in modern acoustofluidics violate a subtle but fundamental Faraday wave theory assumption: the excitation and response frequencies must be similar in magnitude~\cite{perlin2000capillary,binks1997nonlinear}. Curiously, there have been many reports of capillary waves arising in systems where the Faraday wave theory \emph{cannot} apply \cite{kurosawa1997characteristics,lang1962ultrasonic,Blamey:uq}. For example, in a 1~$\mu$L sessile water droplet, visible capillary waves at the droplet's natural frequency ($\mathcal{O}[10^{2}\text{ Hz}]$) arise from acoustic waves at $\mathcal{O}[10^{7}\text{ Hz}]$ or more, five or more orders of magnitude greater in frequency \cite{Blamey:uq}. Remarkably, there are no appropriate theories to predict capillary wave generation nor atomization in these systems. The mechanism of energy transfer across these vastly disparate scales remains unresolved.

    Furthermore, an important traditional assumption made in theoretical studies of a droplet's oscillation is that the perturbation of the fluid interface from the static shape is infinitesimally small~\cite{strani1984free}. This is acceptable for low frequency, low power acoustic waves since their wavelengths are much larger than the droplet's characteristic length scale, producing locally small distortions in the interface. This approach is inappropriate for droplets excited by high-frequency acoustic waves. When the acoustic wavelength is equal to or smaller than the radius of a droplet exposed to the acoustics, pressure nodes and antinodes will be produced along the fluid interface, causing it to significantly deform into a static shape dependent upon the location of these nodes and antinodes \cite{suryanarayana1991effect}. This static finite deformation has notably been observed in the study by Manor \emph{et al.} of a 2~$\mu\ell$ droplet atop a lead zirconate titanate thickness-polarized disk transducer operating at 2~MHz~\cite{manor2011substrate}. 

\begin{figure*}[!t]
            \centering
            \includegraphics[width=0.8\textwidth]{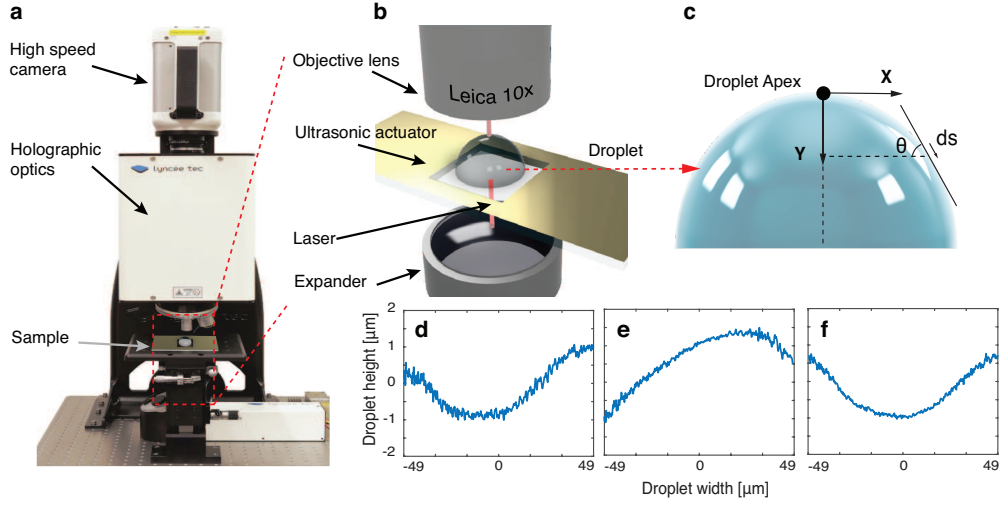}
            \caption{(a) Experimental setup with the DHM and thickness mode device, with a high-speed camera atop the holographic interference optics. Laser light is transmitted to a splitter below the sample plane, with some of the light transmitted as reference light directly into the holographic  optics and (b) some of the light propagated through a laser expander before passing through the transparent acoustic device and the fluid droplet sample before propagating onwards into the objective lens to interfere with the reference light and produce a hologram at the high-speed camera's aperture. Changes in the media along the path of the laser light shift the light's phase with respect to the reference light, causing interference. The (c) droplet itself is shown as a schematic with coordinate definitions. Typical (d-f) periodic distributions of the droplet's deformation appear near its apex (within 49~$\mu$m of the apex (c) along the $x$ axis). Here the deformations are shown at 1.95~ms, 6.55~ms, and 10.92~ms, showing capillary waves appearing due to 6.6~MHz acoustic energy at an amplitude of 1.4~nm being passed from the substrate into the droplet.}
            \label{experiment}
        \end{figure*}

    In this paper, we report the use of high-speed digital holography to measure acoustically-driven, micro-scale capillary waves on a droplet's surface at unprecedented spatiotemporal resolutions. While traditional methods based on high-speed digital photography typically provide $\sim 1 \mu$m displacement accuracy, our digital holographic microscope (DHM) provides measurements to $\mathcal{O}(10^{-9})$~m displacement accuracy normal to the fluid interface, with frame rates up to $116$~kHz and 4~megapixel images for the entire field of view. This provides voluminous information on the dynamic shape of the fluid interface. We then employed particle tracking techniques to observe the pressure distribution and flow pattern in the droplet. This led to our hypothesis that the capillary wave is driven by a feedback mechanism between the acoustic radiation pressure distribution and the droplet's interface with air (\emph{see} Fig.~\ref{experiment}). To test the hypothesis, we first created a physical model of the posited feedback mechanism, mimicking the energy transfer from high-frequency (MHz and beyond) ultrasound to low-frequency capillary waves that appeared upon the droplet surface. We then compared the results produced from simulations using this model with our experimental DHM data collected using droplets of different fluids. Finally, a non-dimensional analysis was derived from the physical model to produce a collapse of measurement data from the water-glycerol system, supporting our model and hypothesis in interpreting the peculiar behavior of capillary waves generated from incident ultrasound.

     
    
    

\section{Physical models and simulation methods}
\subsection{Formation of an acoustic pressure cavity from the droplet: simulation\label{cavity}}
        
         Estimating via Stokes' law \cite{Stokes}, the attenuation length of 1--10~MHz acoustic waves in liquids is generally much larger than the radius of the droplets ($\sim\!1$~mm) under consideration in our system. In this situation, an acoustic wave passed into a droplet would propagate to the opposite side and reflect back \cite{Reflection} from the interface multiple times to form a three-dimensional standing wave in the droplet. The droplet forms an acoustic cavity bound on one side by the solid substrate and the other by air, in each case representing a significant acoustic impedance change that produces the internal reflections in the cavity. 
         
         Here, we employ the standard mass and momentum conservation equations~\cite{nyborg1965acoustic,riaud2017influence},  $\frac{\partial\rho}{\partial t}+\nabla\cdot(\rho u)=0$ and $\rho\frac{\partial u}{\partial t}+\rho(u\cdot\nabla)u=-\nabla p+\mu\nabla^2u+(\mu_B+\frac{\mu}{3})\nabla\nabla\cdot u)$, respectively, where $\rho$ is the fluid density, $u$ is the fluid velocity, $P$ is the fluid pressure, and $\mu$ and $\mu_B$ represent the shear and bulk viscosity, respectively. We are interested in the onset of capillary waves on a droplet's fluid-air interface from acoustic energy introduced into the parent droplet. Since the vibrational velocity from the ultrasonic device required to initiate the capillary wave in our study is small, the so-called \emph{slow streaming} assumption \cite{Friend:2011ss} may be used in the analysis; the basic derivation procedure is provided in section~\ref{slowstreaming} of the Supplementary Information. The key consequence is the ability to decompose the conservation equations~\cite{hunt1955notes,nyborg1965acoustic,Friend:2011ss} into equations that separately represent the fluid dynamics without consideration of the acoustic wave (the zeroth-order domain, denoted later with a ``0'' subscript), the acoustic wave dynamics (the first-order domain, denoted later with a ``1'' subscript), and the consequent acoustic streaming effects that both appear at a higher order and which we ignore based on experimental evidence provided later. The equations can be further simplified~\cite{zarembo1971acoustic} to produce
         \begin{equation}
         \frac{\partial \rho_1}{\partial t}+\rho_0(\nabla\cdot u_1) = 0 \text{ and } \rho_0\frac{\partial u_1}{\partial t} = -\nabla p_1. 
         \label{eqn:pressurewaveeqns}
         \end{equation}
         Together with the linear approximation to the equation of state, $p_1 = c_0^2\rho_1$, the linear pressure wave equations~\eqref{eqn:pressurewaveeqns} can be used to describe the acoustic wave in the fluid with small Mach and Reynolds numbers. We next solve for the radiation pressure using eqns.~\eqref{eqn:pressurewaveeqns}. For boundary conditions, we employed an acoustic impedance-based boundary condition at the fluid-air interface, with the associated acoustic impedance calculated from the standard properties of air and the fluid \cite{jones1997comparison, rezk2014poloidal}. Vital to the analysis is the viscous attenuation of the acoustic wave as it propagates within the droplet; this is equivalent to exponential decay in the acoustic pressure of the wave along its propagation path. The attenuation factors are calculated from the properties of the fluid and the acoustic waves \cite{takamura2012physical}. The attenuation factor for acoustic waves can be expressed as $2\left(\frac{\alpha V}{\omega}\right)^2=\frac{1}{\sqrt{1+\omega^2\tau^2}}-\frac{1}{1+\omega^2\tau^2}$~\cite{morse1986theoretical}, where $\alpha$ is the attenuation coefficient, $V$ is the sound velocity, $\omega$ is the angular frequency, and the relaxation time is $\tau = \frac{4\mu/3+\mu_b}{\rho v^2}$. To accommodate the complex geometries that arise from a finite amplitude deformable fluid interface, we use the finite element method  (COMSOL Multiphysics 6.0, COMSOL, Stockholm, Sweden) in the frequency domain to obtain the pressure distribution. An impedance boundary condition is used to simulate the reflection of the acoustic wave on the fluid-air interface. 
         
\subsection{The droplet's interface shape, defined in part by the acoustic pressure distribution within}
    
        Manor \textit{et al.}~\cite{manor2011substrate} have reported that the acoustic radiation pressure on an air-water interface generated by 2~MHz acoustic waves could cause the droplet to (pseudo)statically deform. In that system and ours, the pressure jump at the interface, surface curvature, and consequent interfacial shape are related to each other according to the Young-Laplace equation. In their system, they assumed the interface remained static; we relax this condition. 
        
        We here assume an axisymmetric droplet shape~\cite{del1997axisymmetric} to conduct a global optimization on the droplet shape subject to volume ($V_0$) conservation and a fixed contact line length ($l_0$) constraint. At large amplitudes, droplet transport is certainly possible \cite{Tan:2007lr}, but here we constrain ourselves to the case where the droplet remains pinned, an assumption made based upon observations of many droplets in our experiments. Thus, the classical Laplace equation can be expressed as a function of the arclength, $s$, of the interface and contact angle, $\theta$:
        \begin{equation}
            \frac{d\theta}{ds} = 2b+cz-\frac{\sin\theta}{x}+\frac{P_a}{\gamma},
            \label{laplace}
        \end{equation}
        with $\frac{dx}{ds} = \cos\theta$, $\frac{dy}{ds} = \sin\theta$ and $\frac{dV}{ds} = \pi x^2\sin\theta$. The constant $c$ is the gravity constant, and $b$ is the curvature of the droplet at its highest point, which is treated as an additional variable to solve the equation with a Neumann boundary condition (${d\theta}/{ds}=b$ at $s=0$). The problem is simplified into a two-dimensional case based on the axisymmetric assumption; $x$, $y$ and $dV$ represent the position and the differential volume at the corresponding position in this axisymmetric system. The purpose of the analysis in subsection~\ref{cavity} was to provide the acoustic pressure $p_a$ present upon the interface that is needed to bring closure to eqn.~\eqref{laplace}. Solving the equation will then produce a deformed surface shape. However, the interface shape produced from this solution changes the shape of the droplet, or, more correctly in this context, the \emph{acoustic cavity}. This will cause a change in the distribution of the standing acoustic wave field in the droplet-based acoustic cavity, leading to a change in the interfacial shape, and so on. 
        
        This \emph{pressure-interface feedback model} mimics the feedback loop that we hypothesize is actually present between the acoustic pressure distribution and the shape of the droplet's interface. 
        
    \subsection{Global optimization to solve the pressure-interface feedback model}
    
        Since the characteristic time of the acoustic wave propagation in a $\mu$L-size droplet ($\sim 10^{-7}$s) is much shorter than that of the capillary wave dynamics ($\sim 10^{-3} $s), the acoustic waves are expected to be reflected multiple times, stabilized, and form compressed and rarefied regions. To study the interaction between the acoustic pressure distribution and the shape change of the droplet, we created a pressure-interface feedback model decoupling the acoustic pressure distribution stabilization and interface shape change processes, assuming the acoustic pressure distribution state is quasi-static. 
        
        In our model, the timing of the changes in the interface from one quasi-static state to the next relies on the classic capillary wave dispersion relation. The time between two simulated states must be estimated, without the dynamic expressions from the fluid mechanics that would be necessary to produce a prediction of the droplet's changing shape over time. In theory, we could simply use direct numerical simulation of the droplet and its response to the acoustic wave. However, direct analysis of the droplet's behavior would be prohibitively expensive given the widely different spatiotemporal scales between the acoustics and hydrodynamics, an issue discussed at length in Orosco and Friend \cite{orosco2021modeling}. Here, we approximate the time interval between each state of the droplet system using the capillary wave dispersion relation, $\omega = \sqrt{\frac{k^3\sigma}{\rho}\tanh{(kh)}}$. For our system, spatial fast Fourier transform analysis was used on the computed profile of the droplet interface to identify the maximum response of the interface at wave number $k=4061$~m$^{-1}$. Based on the dispersion relationship, the frequency of the computed droplet vibration was then approximately 350~Hz, which will be later shown to be of the same order as the experimentally-observed droplet vibration frequency. 
        
        One other challenge is that the curvature $b$ of the wave remains unknown without measurement in a specific system: we have no idea what the value should be in order to conserve the fluid volume as the interface deforms. We overcome this issue by using the shooting method \cite{rapacchietta1977force} on eqn.~\eqref{laplace} and its constraints to arrive at an optimal value of $b$ that conserves the fluid volume ($\sum_i dV_i = V_0$). The pressure-interface feedback model was implemented in a doubly-looped analysis as shown in Supplementary Fig.~\ref{algorithm}, with the outer loop shooting values of curvature, $b$, on the droplet's apex and the inner loop incrementally solving the Young-Laplace equation~\eqref{laplace}. The curvature is first guessed at the apex. This is used to compute the curvature at a fixed point in time progressively across the rest of the fluid interface based on the acoustic pressure distribution and surface tension present at that moment. The shape of the droplet is calculated with the fixed contact length assumption until the last pressure data point is reached. Based on this shape of the overall droplet interface, we then compute the droplet volume and compare this to the conserved value expected from previous steps. The curvature is then shot again based on this result to improve it until the volume is conserved. After optimizing the surface shape of the droplet, the interface is then updated and imported back for simulating the acoustic pressure distribution for the next quasi-static state.
        
\section{Results and Discussion}

To clarify how the energy is transferred from the ultrasonic device's vibrations to interfacial capillary waves, we conducted particle image velocimetry (PIV) experiments with a high-speed camera (FASTCAM MINI, Photron, San Diego, CA USA) and a randomized dispersion of $3$~$\mu$m diameter fluorescent polystyrene particles (Fluoresbrite YG Microspheres, Polysciences, PA, USA). The size was selected to be much smaller than the wavelength of the progressive acoustic wave in the fluid bulk, leaving the viscous drag from the fluid flow to dominate their motion over any acoustic radiation pressure forcing~\cite{dentry2014frequency,orosco2021modeling}. 

A key mechanism responsible for particle motion in this system could be acoustic streaming. The associated energy transfer from the underlying acoustic wave in the substrate to fluid flow could perhaps generate capillary waves, as posited in past work \cite{Blamey:uq} and seen and used in many other sessile droplet experiments \cite{Rufo:2022vq}. Acoustic streaming \cite{Friend:2011ss} is generated by a nonlinear interaction between an acoustic wave and the medium it is propagating through~\cite{lighthill1978acoustic,Orosco:2022vk}, and may arise at the boundary \cite{schlichting1932berechnung} or in the bulk of the fluid \cite{eckart48}. It is commonly seen when the frequency and amplitude of the ultrasound are high, where a greater proportion of the acoustic energy is transferred into net fluid flow \cite{dentry2014frequency,Orosco:2022vk}. The induced flow, especially the flow immediately beneath the fluid interface, could give rise to capillary waves through a type of viscous Kelvin-Helmholtz instability \cite{hooper_boyd_1983}, thoroughly explored in the context of Faraday waves by Vega et al. \cite{Vega2004}.

\begin{figure}
 \includegraphics[width=0.8\textwidth]{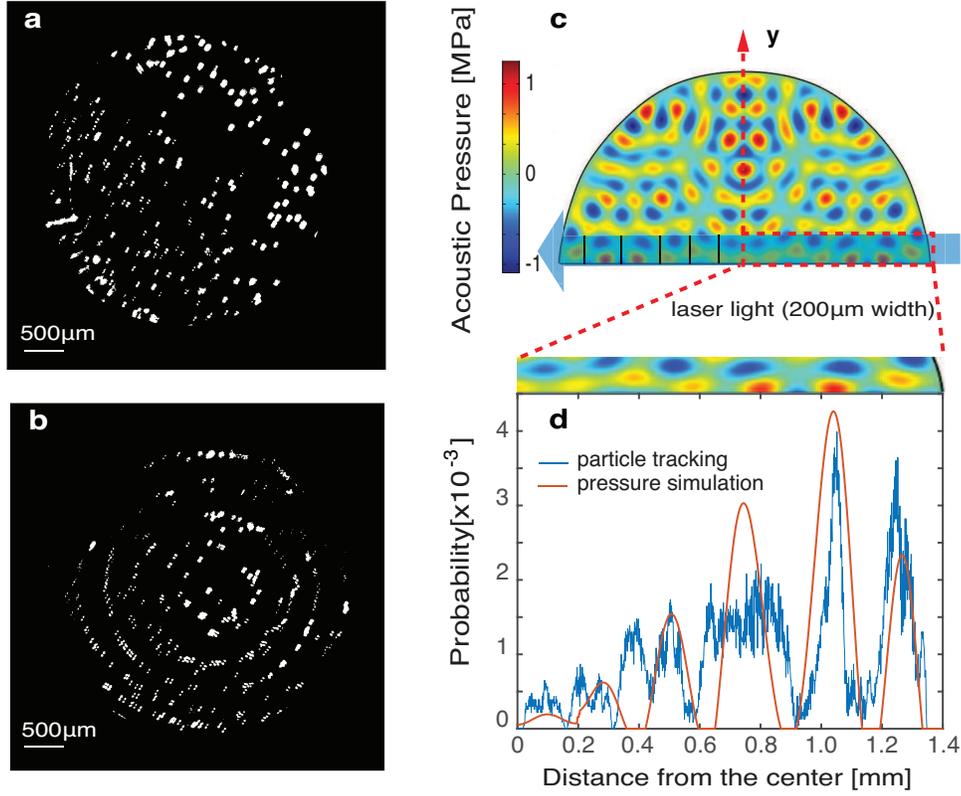}
 \caption{Particle tracking images extracted from high-speed video: (a) before and (b) after applying the 6.6~MHz acoustic excitation with 1.5~nm amplitude for 0.48~s. The particles The ring-shaped pattern forms as particles migrate to pressure nodes formed by acoustic excitation. The (c) computed pressure distribution within the droplet shows a complex but consistent standing wave. (d) The (red line) computed pressure field generated by this system at the fluid-substrate interface shows close correspondence with (blue line) the probability of entrapped particles' positions obtained from particle tracking experiments.\label{fp}}    
\end{figure}

We examined the region near the substrate where induced flow through acoustic streaming would be especially evident, indicated in Fig.\,\ref{fp}(c) with a right-to-left blue arrow. Figure~\ref{fp}(a,b) shows the distribution of the particles before and after a period of time after applying 6.6~MHz acoustic excitation at an amplitude of 1.5~nm to a 5~$\mu$L DI water droplet, respectively. In Fig.\,\ref{fp}(a), before the application of the acoustic wave, the particles were randomly distributed in the droplet. After applying the acoustic wave for 0.48 seconds, the particles migrated to well-defined positions forming a ring-like pattern in Fig.\,\ref{fp}(b). The migration distances from the particles' original positions to the neighboring pressure nodes were short ($\sim 10^{-7}$~m). For the input amplitudes used in this study, the fluid bulk was observed to be essentially quiescent in the PIV experiment. The slow motion of the particles further convinced us that it was the acoustic pressure instead of the streaming that dominated the system when the capillary waves were initiated, considering how weak this effect is on the particles \cite{Li:2007fk}. The capillary waves we observed are, therefore, not the result of acoustic streaming or other induced flow behaviors.

The results of our simulation were confirmed with experimental particle migration measurements. The numbers of particles with different distances from the droplet's center in Fig.\,\ref{fp}(b) were counted, and the corresponding probability density distribution was plotted  in Fig.\,\ref{fp}(d: blue curve). The results of the acoustic pressure simulations are shown in Fig.\,\ref{fp}(c: red curve). A complex field of positive and negative pressure nodes is formed within the droplet. Within a stable oscillating pressure field, particles are driven from positive pressure nodes to the closest positions with negative acoustic pressure. To compare the experimental data to the simulated position of the pressure nodes, we take the average of the pressure simulated in different layers along the y-axis at the bottom of the droplet (blue region shown in Fig.\,\ref{fp}(c)). Since the particles migrate toward the closest negative nodes, the probability associated with a particle migrating to a given position is proportional to (i) the pressure, and (ii) the number of particles in the region. We divide the illuminated area into several regions according to the midpoints between any two neighboring negative pressure nodes (black lines through the midpoints in Fig.\,\ref{fp}(c)). 
The red curve in Fig.\,\ref{fp}(d) represents the normalized probability corresponding to migrated particle positions based on the simulated pressure results. The data collected from the particle tracking experiments find good agreement with the magnitude, number, and location of the pressure nodes predicted by the model. Since acoustic streaming effects are negligible within the droplet, these results provide strong evidence for the existence of a stably oscillating, spatially localized pressure distribution. It can be seen here that with high-frequency ultrasound, the acoustic wavelength is on the order of, or smaller than, the size of the droplet. When properly accounted for, the effects of reflection and attenuation of the acoustic waves and their interactions serve to redistribute pressure within the droplet in a manner that is highly consistent with our observations. This demonstrates a clear, intuitive mechanism for the noted energy transfer across wavenumbers spanning many orders of magnitude, a mechanism that is quite different from the mechanisms proposed by using classical theory.

\begin{figure*}[t!]
        \centering
        \includegraphics[width=0.9\textwidth]{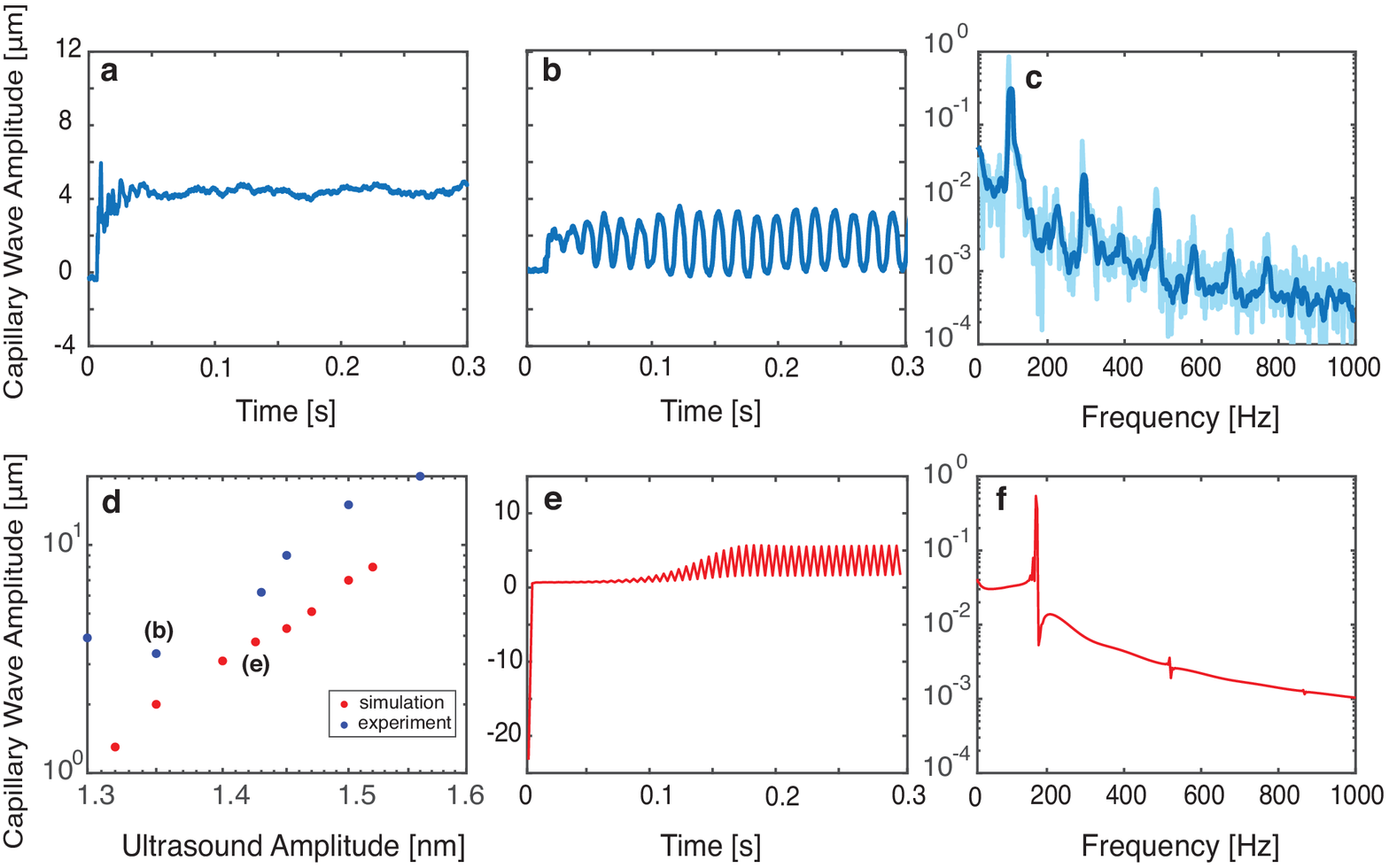}
        \caption{Vibration at the apex of a 5~$\mu$L water droplet collected with the DHM system during application of acoustic forcing. (a) The droplet's shape stabilizes after an initial step change of about 4~$\mu$m upon excitation from a 1.1~nm amplitude acoustic wave. (b) With an increase of the acoustic wave amplitude to 1.35~nm, capillary waves are generated on the interface. The (e) results of a corresponding simulation of capillary waves induced by $1.43$~nm acoustic waves shows a phenomenological similarity though with a higher oscillation frequency. Nevertheless, the (d) capillary wave amplitudes from (blue) experiments and (red) simulations closely correspond; the data points corresponding to the (b,e) capillary wave oscillations are marked. Moreover, the FFT spectra corresponding to the (b,e) time domain plots are given in (c) for the experimental results (light blue: raw data, dark blue: smoothed data) and (f) for the computational results, respectively.}
        \label{f2}
\end{figure*}%

We then observed the vibration of the droplet with the DHM system. We first examined the effect of increasing the input vibration amplitude upon the onset and growth of the capillary wave at the fluid interface. %
The response of the droplet's apex, in particular, is shown in Fig.~\ref{f2}. With the droplet and experimental setup intact, the vibration amplitude was controlled by measuring it using a laser Doppler vibrometer (UHF-120SV, Polytec, Waldbronn, Germany) while adjusting the signal input. %

\begin{figure}[ht]
\includegraphics[width=0.75\textwidth]{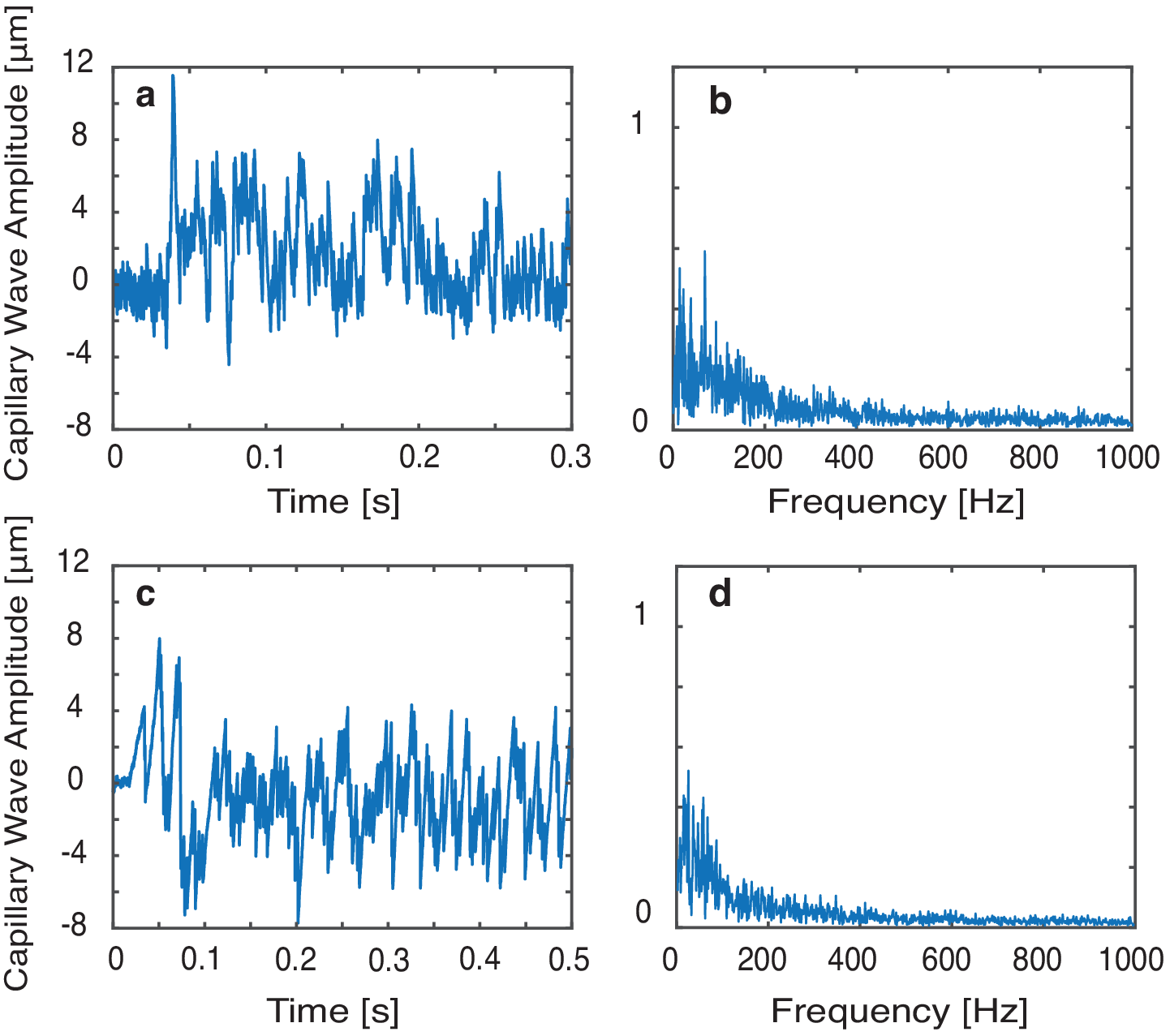}
\caption{Increasing the input 6.6~MHz acoustic amplitude to (a,b) 1.7~nm with a 5~$\mu$L water droplet and (c,d) a 90\%-10\%wt glycerol water mixture at an input amplitude of 5.1~nm  produces distinctly nonlinear capillary wave responses. The (b,d) FFT results of the resonance peaks (vibration modes) are sufficiently broadened by the nonlinear response as to lose their peak-like character (\emph{compare these with} Fig.~3(c,f)). This is seen in the time domain as (a,c) a seemingly inharmonic capillary wave response.}
\label{nonlinear}
\end{figure}%

The experiments revealed three regimes. In order of increasing input amplitude, these are a static shape change (Fig.~\ref{f2}(a) and section~\ref{staticmode} in the Supplementary Information), akin to past observations; steady vibration (Fig.~\ref{f2}(b)); and nonlinear vibration (Fig.~\ref{nonlinear}). Care was taken to isolate the system from ambient vibration and air currents in the laboratory. 

A sudden, static change of the droplet height was observed at the moment acoustic excitation of amplitude $\leq 1.3$~nm was applied (Fig.\,\ref{f2} (a), at 0.005~s). This occurs due to a sudden change in the pressure at the interface resulting from acoustic radiation forces upon it from below. Interestingly, the lack of oscillatory motion of any kind indicates that, by itself, the acoustic wave propagating through the fluid and reflecting from the interface is insufficient to produce capillary waves. This suggests the existence of another mechanism facilitating the energy transfer from the incident acoustic wave to the generation of capillary waves.
  
Increasing the acoustic excitation amplitude to more than $1.3$~nm produces capillary wave oscillations. A spontaneous shape change is still observed when the input signal is initiated. Following this, the droplet interface grows to exhibit a capillary wave oscillation that becomes stable over time. Figure\,\ref{f2}(b) is an example of this response from a 1.35~nm amplitude input. The corresponding frequency response is provided in Fig.\,\ref{f2}(c), showing several resonance peaks within the range 0 to 800~Hz. This response can be placed into the context of Rayleigh's equation, which predicts the resonance frequencies of a droplet's surface based upon its volume and density while neglecting air that surrounds it \cite{bostwick2009capillary}:
        \begin{equation}
            f = \frac{1}{2\pi}\sqrt{\frac{l(l+1)(l+3)\gamma}{\rho R^3}},
        \end{equation}
where $R$ is the radius of the droplet, $\gamma$ is the interfacial surface tension, and $l=1,2,\ldots$ is the mode number. In our system here with a 5~$\mu$L water droplet, the first three natural frequencies are predicted to be $78.17$~Hz, $151.39$~Hz, and $234.53$~Hz. The first frequency predicted with Rayleigh's equation roughly corresponds to the first observed resonance peak (96.85~Hz) in Fig.\,\ref{f2}(c). Given the many simplifying assumptions in Rayleigh's equation, it is remarkable that a sessile droplet's oscillatory response reasonably compares to it, an indirect indicator of the relatively weak influence of the pinned boundary and configuration on the response.

As the input acoustic amplitude continues to be increased, nonlinearity plays a larger role in the capillary wave dynamics. Evidence of this is provided in Fig.\,\ref{nonlinear}(a,b). In Fig.\,\ref{nonlinear}(a), the wave pattern is nonuniform, and no obvious period of oscillation can be directly observed. The narrow resonance peaks observed in the stable capillary wave oscillations in Fig.\,\ref{f2}(c) are broadened to essentially eliminate the peaks in Fig.~\ref{nonlinear}(b), due to the non-resonant interaction between capillary waves of different frequencies that give rise to new capillary waves. These interactions generate waves with wavelengths, $\lambda$, and frequencies, $f$, obeying a more generalized dispersion law \cite{orosco2022identification} than those derived from linear theory, such as $\omega^2 = \left(2\pi f\right)^2 = \frac{4\pi^2\gamma}{\rho\lambda^3}$ from Lamb \cite{lamb1993hydrodynamics}. The distinct change in the frequency response is the principal means to distinguish steady-state vibrations from nonlinear oscillations in this system. In the absence of significant nonlinearity in the system, a wave that is incongruous with the resonant response of the fluid interface---possessing a different frequency or wavelength than one of the admissible waves---will vanish. In a capillary wave system with nonlinearity, however, the nonlinearity acts to broaden the resonant responses. Each resonance is broadened by the nonlinearity to a spectral range; this \emph{nonlinear broadening} of the dispersion relationship permits newly generated waves in this broader range to persist~\cite{berhanu2018turbulence,kartashova2010nonlinear}.

        \begin{figure*}[!t]
            \centering
            \includegraphics[width=0.8\textwidth]{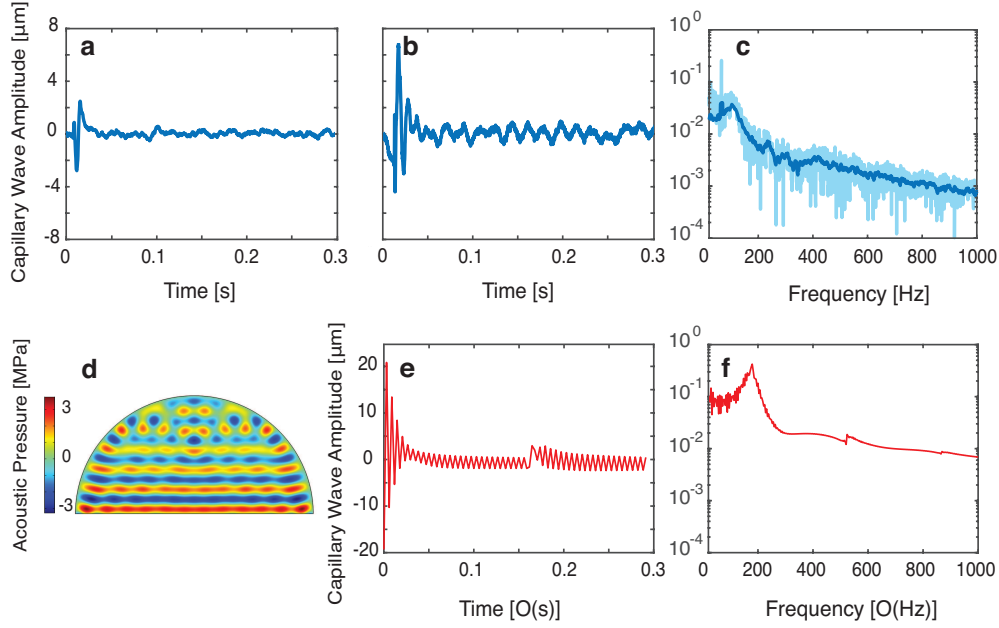}
            \caption{Vibration pattern of a 90\%-10\% wt glycerol-water solution droplet collected with the DHM system before and during application of acoustic forcing. The  (a) droplet's vibration attenuates after approximately 0.025~s of vibration excitation at $3.4$~nm amplitude. (b) Linear and stable oscillation is observed in the same system while using a higher, $3.9$~nm, input acoustic wave amplitude after an initial transient and exponential decay of a larger wave at the start of excitation. The (c) FFT-derived spectral content of this response shows some key peaks below 200~Hz (light blue: raw data, dark blue: smoothed data). (d) Simulation of the capillary wave phenomena as driven from a 3.9~nm input shows (e) a similar response though at a higher frequency. The acoustic wave forms a standing wave in the droplet, with (d) weakening amplitude near the top of the droplet in part due to the attenuation in the highly viscous fluid. The (f) FFT spectrum of the computational result resembles the (c) experimental result.}
            \label{f5}
        \end{figure*}

    In order to analyze the dynamic droplet shape change induced by acoustic pressure feedback, we have developed a pressure-interface feedback model by extracting the simulated pressure data from the surface of the droplet and utilizing the data to compute an update to the modified Young-Laplace boundary condition (eqn.\,\eqref{laplace}). Here, the surface tension balances the acoustically-driven dynamic pressure jump by inducing local curvature. The direction of the change is determined by the sign of the local pressure change. At each step in the simulation, the shape that is deduced by optimizing the curvature against the Young-Laplace boundary is then utilized to compute an updated pressure distribution. This update is then used in turn, along with the Young-Laplace condition to update the droplet shape. Iterating accordingly, we obtain a time series of states of the droplet shape and pressure distribution. The comparison of experimental and simulation results of capillary waves with steady vibration states are shown in Fig.\,\ref{f2}(d). Figure~\ref{f2}(e) shows one simulated case for a small input amplitude, $1.43$~nm. The droplet experiences a nearly instantaneous height change when the input is switched on and followed by stable capillary oscillations with amplitudes of around $3.7$~$\mu$m. This directly corresponds with experimental observations of the linear vibration mode. Since the oscillations are linear, we can correlate the inter-frame time scale $\Delta t$ using a simple oscillator model to show that the simulated oscillation of the droplet is in the low-frequency range observed in the experiments. The FFT comparison in Fig.\,\ref{f2}(c,f)) reveals a capillary wave spectrum similarity between the experimental measurements and the pressure-feedback simulation.


        \begin{figure}[!htb]
            \includegraphics[width=0.7\textwidth]{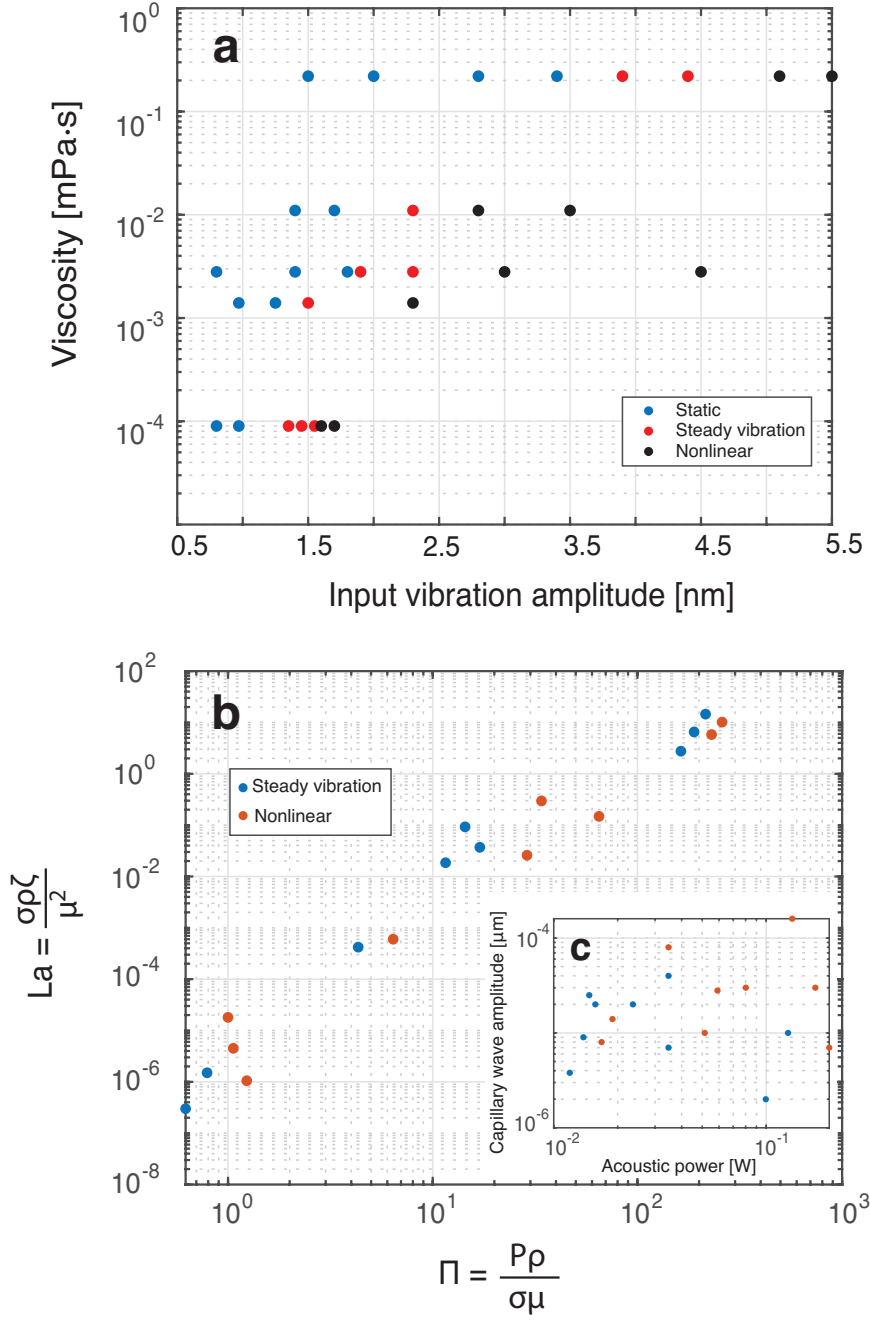}
            \caption{(a) The distribution of the capillary waves' status observed in experiments with different liquid viscosity and acoustic wave amplitude. (b) The experimental data with steady vibration mode and nonlinear mode collapses to similarity within the nondimensional basis noted along the axes, while there is no pattern that can be found with the data plotted in the inset (c) before nondimensionalization.}
            \label{nondi}
        \end{figure}

Weak attenuation is an important factor that affects the complexity of the pressure distribution. With an attenuation length $1/\alpha=0.034$~m in water~\cite{Attlength}, acoustic waves are reflected at the boundaries multiple times before fully attenuating within the millimeter-sized droplet. To study attenuation effects on capillary wave formation, we conducted experiments and simulations for a 90\%-10\% glycerol-water solution. We use glycerol since it has a similar density (1260~kg/m$^3$) and surface tension (63.4~mN/m) to water~\cite{takamura2012physical}, but a substantially higher viscosity, leading to an attenuation distance that is roughly one order of magnitude smaller than water: $1/\alpha = 2.8\times10^{-4}$~m. This allows us to isolate the effect of attenuation on capillary wave formation. 

The results for the solution are similar to those for water. A static shape change (Fig.\,\ref{f5}(a)), steady vibration (Fig.\,\ref{f5}(b)) and nonlinear vibration mode (Fig.\,\ref{nonlinear}(c)) are also observed in the glycerol-water solution droplet. Compared to the vibration of the water droplet, the droplet height tends to overshoot the static displacement to a much greater extent for the more viscous fluid after the initial onset of acoustic power. These large amplitude vibrationsstill, however, exponentially decay to either a static displacement for weak excitation (3.4~nm in Fig.\,\ref{f5}(a)) or uniform oscillations for larger amplitude excitation ((3.9~nm in Fig.\,\ref{f5}(b)). The exponential decay can be observed in both the experimental and simulation results (Fig.\,\ref{f5}(b) and (e)).  

Simulations were conducted with the same parameters used in the experiments and the results for the $3.9$~nm input amplitude are shown in Fig.\,\ref{f5}(b,e). Figure~\ref{f5}(d) shows the acoustic pressure distribution in the droplet. A laminar pressure distribution was observed with nodal formation near the top portion of the droplet. The input amplitude threshold for capillary wave generation was confirmed with the experiment, as shown in Fig.\,\ref{f5}(b), providing further evidence in support of the pressure feedback model. Compared with the capillary wave vibration pattern on the water droplet surface, there exists a more obvious amplitude decay in both the simulation and the experiment after the acoustic wave is initiated.

Looking more broadly, these capillary wave states exist for specific choices of viscosity and input vibration amplitude. A map of this is provided in Fig.\,\ref{nondi}(a). As the viscosity increases, the input amplitude likewise must increase to produce similar wave states. The steady vibration state itself is present as a narrow region between the static deformation and nonlinear vibration states on the map. 
    
Next, we consider whether these wave states can be described in a nondimensional representation. Using dimensional analysis, we identified the Laplace number $\text{La}=\frac{\sigma\rho\zeta}{\mu^2}$ as the relevant nondimensional number to describe the capillary wave behavior; $\zeta$ is the capillary wave amplitude. The Laplace number relates the conservative surface tension forces to dissipative viscous forces. If the Laplace number characterizes the system output (i.e., capillary waves), then the dimensionless number characterizing the input becomes $\Pi = \frac{P\rho}{\sigma\mu}$, where $P$ is the input acoustic power. Applying this non-dimensionalization produces Fig.\,\ref{nondi}(b), which collapses the source data plotted as an inset in Fig.\,\ref{nondi}(c). Though it does not separate the steady and nonlinear vibration wave behavior, it does suggest that, for a known fluid, a power law relationship between La and $\Pi$ that approximately describes the capillary wave amplitude.

\section{Conclusions}
A new method to observe the onset and growth of capillary wave motion on fluid interfaces from high-frequency acoustic waves has been provided using high-speed digital holographic microscopy. The results produced from this method are compared to a new approach to the solution of capillary wave dynamics through the use of a hybrid solution method. This method employs a two-step process, first producing the pressure distribution on the fluid interface from the relatively fast acoustic standing wave distribution in the acoustic cavity formed by the droplet. This step is followed by a computation of the new shape of the fluid interface that would arise as a consequence of the new pressure distribution taking into account the acoustic pressure variation at the interface. Thus, the model is built crucially upon the assumption of a pressure-interface feedback mechanism governing the onset of capillary waves across several orders of magnitude in spatiotemporal scale disparity. There is good correlation between the computational results produced using this method and the experimental observations. Further refinements of this method are likely to produce improvements in the frequency predictions for the induced capillary waves, and additional physical insight into the complex phenomena of capillary wave generation.

\section{Experimental Methods and Physical Models}
\subsection{Fabrication of high-frequency ultrasonic transducer}
    
The ultrasonic devices were fabricated from $128^{\circ}$ Y-rotated, X-propagating lithium niobate wafers with 500~$\mu$m thickness and mirror-finish polishing on both sides (Roditi, London, UK). On each side of the wafer, the sputter deposition method (Denton Discovery 18, New Jersey, USA) was used to deposit a layer of chromium and a 400~nm layer of gold. These provided electrodes to facilitate the driving of the thickness mode. One 0.5~cm $\times$ 0.5~cm region at the center of each transducer was blocked with sacrificial photoresist to prevent gold deposition, leaving this region transparent for the digital holographic microscope (DHM) laser to pass through  during experiments (Fig.\ref{experiment}(a) and (b)). This region was sufficiently small that the overall displacement profile of the substrate was nearly constant over the transparent and gold-plated regions when used as a transducer.
        
\subsection{Capillary wave generation}

Thickness mode vibrations were induced by applying an amplified voltage potential at a frequency matched to the thickness resonance of the device (6.6~MHz for the 500~$\mu$m thick wafer). A 5 $\mu\ell$ droplet was dispensed onto the center of the transparent window \cite{CapillaryLen} using a measuring pipette (2-20 $\mu L$, Thermo Fisher Scientific, USA). The resonant frequency and voltage-vibrational amplitude correspondence of the transducer were characterized with laser Doppler vibrometry (LDV; UHF-120, Polytec, Germany).
        
\subsection{Digital holographic microscope}

Measuring microscale vibrations on the surface of droplets is challenging due to the size and speed of the dynamics under consideration: $\sim 1$~nm amplitudes and $\sim 1$~$\mu$s time scales. While the LDV is suitable for single-point and scanning measurements of a surface with well-defined periodic vibrations, our high-speed digital holographic transmission microscope (DHM, Lyncee-tec, Lausanne, Switzerland) utilizes holographic imaging methods combined with a high-speed camera (FASTCAM NOVA S12, Photron, San Diego CA) to characterize interfacial dynamics across an entire region of interest on the liquid-air boundary at up to 116~kfps. It provides real-time three-dimensional surface structure data with 3~$\mu$m lateral spatial resolution and 3~nm displacement resolution.
        
\subsection{Details of the setup to track the particles' movement in droplet}
     
To accurately capture the movement of the particles in the droplet, we employed a high-speed camera to record the process and used fluorescent particles to increase the light intensity. The excitation and emission maxima of the fluorescent particles were at 441~nm and 485~nm, respectively. We illuminated the particles with a blue laser sheet generator (M-Series 450~nm wavelength, with Powell lens; Dragon Lasers, Jilin, China). Aa 450-nm long-pass filter (FEL0450, ThorLabs, Newton NJ) was placed in front of the camera to filter out this excitation light, leaving the light emitted from the particles to be collected by the camera. The thickness of the laser sheet was 200~$\mu$m. For the results provided in this paper, the laser sheet was passed through the bottom of the droplet adjacent the solid substrate.

\section{Acknowledgments}

    The work presented here was generously supported by a research grant from the W.M.\ Keck Foundation to J.\ Friend. The authors are also grateful for the substantial technical support by Yves Emery and Tristan Coloumb at Lyncee-tec, and Eric Lawrence, Mario Pineda, Michael Frech, and Jochen Schell among Polytec’s staff in Irvine, CA and Waldbronn, Germany. Fabrication was performed in part at the San Diego Nanotechnology Infrastructure (SDNI) of UCSD, a member of the National Nanotechnology Coordinated Infrastructure, which is supported by the National Science Foundation (Grant ECCS--1542148).

\bibliographystyle{apsrev4-2} 
\bibliography{reference}

\begin{thebibliography}{57}
\bibitem [{Sto()}]{Stokes}%
  \BibitemOpen
  \href@noop {} {}\bibinfo {note} {$\alpha_l^{-1}=\frac{\rho v^3}{4 \pi^2 f^2
  (\frac{4}{3}\mu+\mu_b)}$\, where $\rho$ is the density of the liquid, $v$ is
  sound velocity, $f$ is the frequency, and $\mu$ and $\mu_b$ are the dynamic
  and bulk viscosity, respectively.}\BibitemShut {Stop}%
\bibitem [{Ref()}]{Reflection}%
  \BibitemOpen
  \href@noop {} {}\bibinfo {note} {The acoustic impedance can be calculated by
  the density of the media and speed of sound $Z = \rho c$. The reflection
  coefficient, which describes how much portion of the acoustic power is
  reflected at the interface, can be expressed as $r =
  {\frac{Z_l-Z_g}{Z_l+Z_g}}^2$, where $Z_l$ and $Z_g$ are the acoustic
  impedances of liquid and gas. With water and air used in this example, more
  than 99\% of the acoustic waves carried in water are reflected from the
  air-water interface, implying that most of the vibrational energy will be
  conserved in the droplet.}\BibitemShut {Stop}%
\bibitem [{Att()}]{Attlength}%
  \BibitemOpen
  \href@noop {} {}\bibinfo {note} {For water droplets, the dynamic viscosity
  $\mu$ and bulk viscosity $\mu_b$ are $0.89$~mPa$\cdot$s and
  $0.2$~mPa$\cdot$s, respectively. With acoustic wave at 6.6~MHz, the
  attenuation distance ($1/\alpha$) is therefore 0.034~m,}\BibitemShut
  {NoStop}%
\bibitem [{Cap()}]{CapillaryLen}%
  \BibitemOpen
  \href@noop {} {}\bibinfo {note} {This droplet volume is used so that the
  radius of the droplet is smaller than the capillary length, $l =
  \sqrt{\gamma/\rho g}$, with $\gamma$ as surface tension and where $\rho$ is
  the liquid density. For the media used in this study, capillary forces
  dominate at the droplet surface and the effect of gravity is
  inconsequential.}\BibitemShut {Stop}%
\end{thebibliography}

\appendix
\renewcommand\thefigure{S\arabic{figure}}    
\setcounter{figure}{0}    

\section*{Supplementary Materials}
\subsection{Derivation of linear mass and momentum conservation equations based on the slow streaming assumption\label{slowstreaming}}

\begin{subequations}
    \begin{align}
        \begin{split}
            \frac{\partial\rho}{\partial t}+\nabla\cdot(\rho u)=0
        \end{split}\\
        \begin{split}
            \rho\frac{\partial u}{\partial t}+\rho(u\cdot\nabla)u=-\nabla p+\mu\nabla^2u+(\mu_B+\frac{\mu}{3})\nabla\nabla\cdot u)
        \end{split}
    \end{align}
    \label{NS}
\end{subequations}

The terms in equations \ref{NS} can be decomposed into three contributions:
\begin{subequations}
    \begin{numcases}{}
      u = u_0 +\epsilon u_1 +\epsilon^2u_2 +\mathcal{O}[\epsilon^3]\\
      p = p_0 +\epsilon p_1 +\epsilon^2p_2 +\mathcal{O}[\epsilon^3]\\
      \rho = \rho_0 +\epsilon \rho_1 +\epsilon^2\rho_2 +\mathcal{O}[\epsilon^3];
    \end{numcases} 
    \label{decompose}
\end{subequations}
$u_0$, $p_0$, and $\rho_0$ are hydrostatic terms and those with subscripts 1 and 2 refer to first and second-order perturbations. The variable $\epsilon$ is a Mach number, defined here as the ratio of fluid velocity to the speed of sound ($\epsilon = u_1/c_0$). Since the fluid velocity is small in this system, $\epsilon\ll 1$. Introducing the expansions \ref{decompose}(a-c) into eqns.~\ref{NS}(a) and \ref{NS}(b) and grouping in terms of $\epsilon$, the resulting equations can be separated into three parts according to the zeroth, first, and second order components of the acoustic perturbation. The first-order acoustic perturbation expression represents the behavior of the linear acoustic waves in the fluid. Since the dimensions of the droplet are small and the fluid velocity is likewise small, the Reynolds number in this case is also a small value. So the equations can be simplified as follows:
\begin{equation}
    \frac{\partial \rho_1}{\partial t}+\rho_0(\nabla\cdot u_1) = 0,
    \label{mass2}
\end{equation}

\begin{equation}
    \rho_0\frac{\partial u_1}{\partial t} = -\nabla p_1.
    \label{momentum2}
\end{equation}

\subsection{Static mode simulation results\label{staticmode}}
\begin{figure}[ht]
    \centering
    \includegraphics[width=0.75\textwidth]{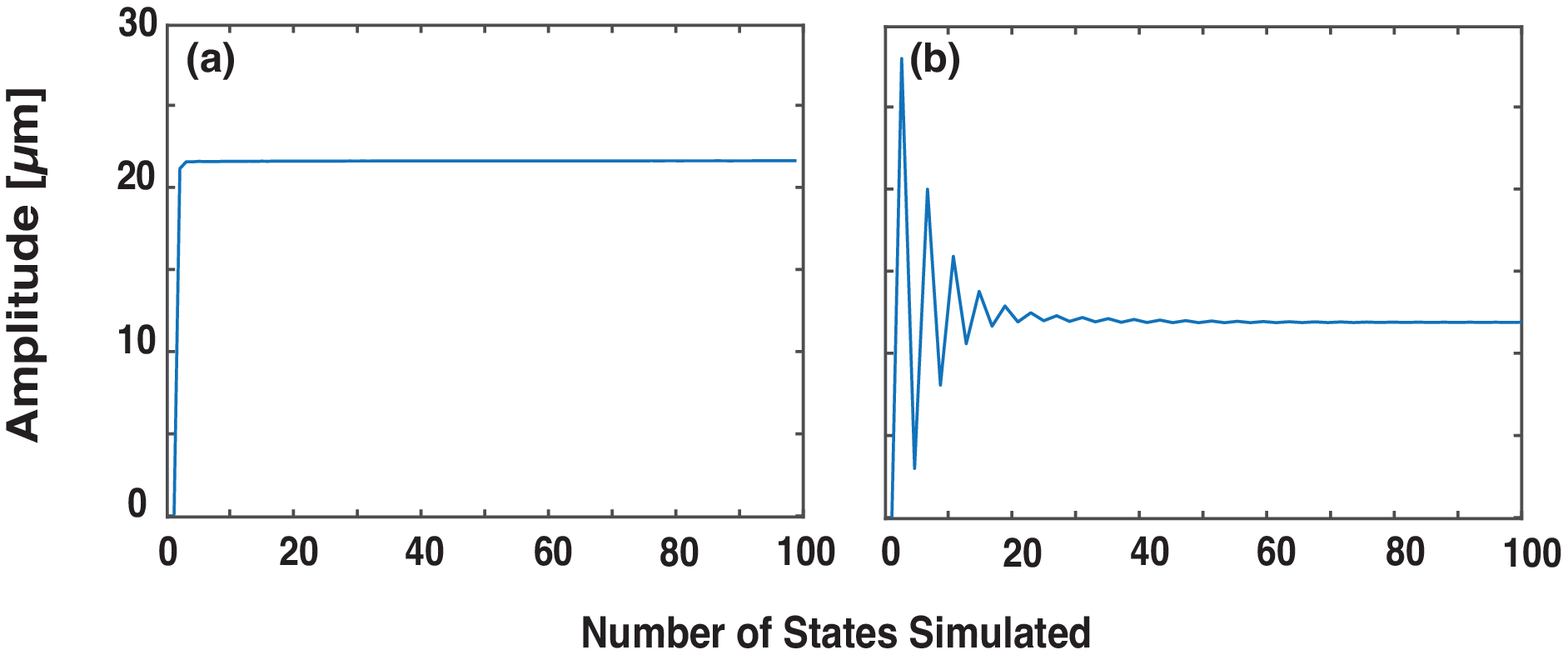}
    \caption{Simulated deformation of the fluid interface in response to 6.6~MHz acoustic waves. A sudden displacement and static response were induced in (a) the water sessile droplet from 1.1~nm acoustic waves, while the (b) 90\%-10\%wt glycerol water droplet surface exhibited both the sudden displacement and the appearance of strongly damped capillary waves from 3.3~nm acoustic waves.}
   \label{ss}
\end{figure}

Two simulation prediction of the static deformation of the fluid interface are provided in Fig.\ref{ss} for (a) water from 1.1~nm acoustic waves, and (b) a 90\%--10\%wt glycerol-water solution from 3.3~nm acoustic waves. After the application of 1.1~nm amplitude 6.6~MHz ultrasound to the water droplet, it produced a sudden increase in height and thereafter a static response, akin to past observations~\cite{manor2011substrate}. Upon the application of 3.3~nm amplitude 6.6~MHz ultrasound to the glycerol-water combination, the sudden increase in height was followed by capillary waves that were strongly damped. The horizontal axis in these results is reported as \emph{number of states simulated} as the simulation is quasi-static. This axis can be transformed to a time-based prediction using the capillary wave dispersion relation as described in the previous subsection. 

\subsection{Algorithm used in the pressure-interface feedback model}
\begin{figure}[ht]
    \includegraphics[width=0.75\textwidth]{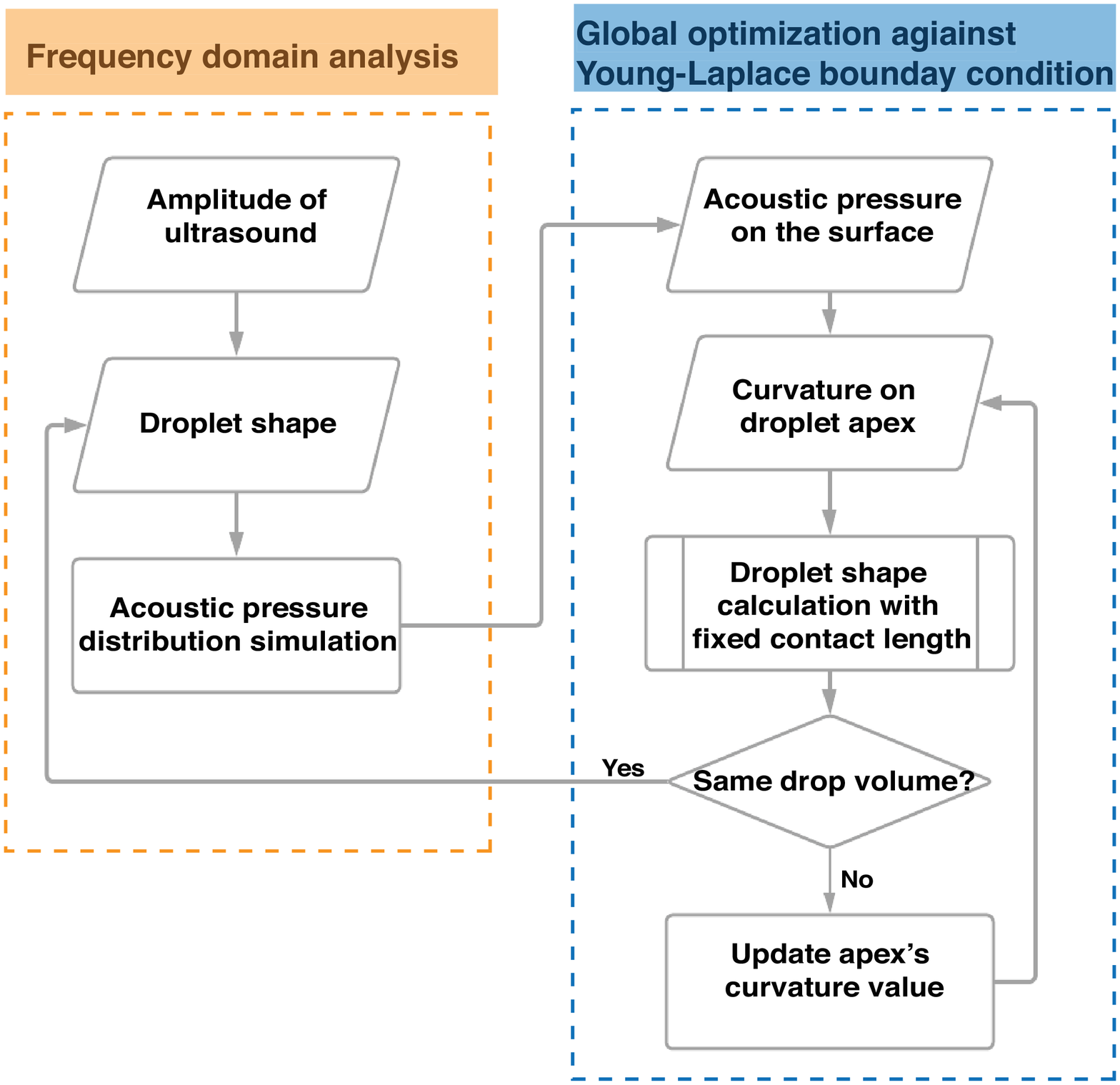}
    \caption{The algorithm to simulate the shape of the droplet and the acoustic pressure distribution}
    \label{algorithm}
\end{figure}
\end{document}